\documentclass[twocolumn, conference, 10pt]{IEEEtran}
\usepackage{amsmath,amssymb,amsthm}
\usepackage{bm}
\usepackage{url}
\usepackage{braket}

\newcommand{\Tr}{\mathrm{Tr}}

\allowdisplaybreaks[4]

\newtheorem{theorem}{Theorem}

\theoremstyle{definition}
\newtheorem{definition}[theorem]{Definition}
\newtheorem{example}[theorem]{Example}

\theoremstyle{remark}
\newtheorem{remark}[theorem]{Remark}

\title{Holographic Transformation, Belief Propagation and Loop Calculus for Generalized Probabilistic Theories}
\author{\IEEEauthorblockN{Ryuhei~Mori}
\IEEEauthorblockA{
Department of Mathematical and Computing Science,
Tokyo Institute of Technology, Tokyo, Japan\\
email: mori@is.titech.ac.jp}}

\begin{document}
\maketitle
\begin{abstract}
The holographic transformation, belief propagation and loop calculus are generalized to problems in generalized probabilistic theories including quantum mechanics.
In this work, the partition function of classical factor graph is represented by an inner product of
two high-dimensional vectors both of which can be decomposed to tensor products of low-dimensional vectors.
On the representation, the holographic transformation is clearly understood by using adjoint linear maps.
Furthermore, on the formulation using inner product, the belief propagation
is naturally defined from the derivation of the loop calculus formula.
As a consequence, the holographic transformation, the belief propagation and the loop calculus are generalized to
 measurement problems in quantum mechanics and generalized probabilistic theories.
\end{abstract}

\section{Introduction}
The computation of the partition function of factor graphs is one of the central problems in
statistical physics, information theory, machine learning and computer science~\cite{mezard2009ipa}.
Recently, a general technique transforming a representation of a partition function into a different representation
is invented, and called the holographic transformation~\cite{valiant2008holographic}, \cite{5695119}.
Many equalities in broad area can be understood by the holographic transformation, e.g.,
the high temperature expansion, the MacWilliams identity, the loop calculus, etc.~\cite{forney2011partition}.
Especially, the loop calculus, which shows the equality relating the partition function and its
approximation obtained by the message-passing algorithm called the belief propagation, is an interesting example~\cite{1742-5468-2006-06-P06009}.
In this paper, the partition function and the holographic transformation are expressed in an abstract way which allows to 
generalize the holographic transformation to problems in quantum information science.
More specifically, the partition function is represented by an inner product on high-dimensional linear space where
each of the two vectors can be decomposed to tensor products of low-dimensional vectors.
On this formulation, the holographic transformation can be clearly understood by using linear maps and their adjoint maps.
This understanding is quite clear and also allows to define the belief propagation
and to prove the loop calculus for the general problem of computing the inner product.
As a consequence, we obtain the holographic transformation, the belief propagation and the loop calculus for measurement problems
 in generalized probabilistic theories including quantum mechanics.
Probability of outcome in measurement-based quantum computation is an important application of this work.

\section{Factor graphs and bipartite normal factor graphs}
A factor graph is a bipartite graph defining a probability measure.
A factor graph consists of variable nodes, factor nodes and edges between a variable node and a factor node.
Let $V$ be the set of variable nodes and $F$ be the set of factor nodes.
Let $E\subseteq V\times F$ be the set of edges.
For a variable node $i\in V$, $\partial i\subseteq F$ denotes the set of neighborhoods of $i$.
In the same way $\partial a\subseteq V$ is defined for $a\in F$.
For each variable node $i\in V$, there is an associated finite alphabet $\mathcal{X}_i$ and an associated function $f_i\colon \mathcal{X}_i\to \mathbb{R}_{> 0}$.
For each factor node $a\in F$, there is an associated function $f_a\colon\prod_{i\in\partial a}\mathcal{X}_i\to\mathbb{R}_{\ge 0}$.
Let $\bm{x}_{V'}\in\prod_{i\in V'}\mathcal{X}_i$ be variables corresponding to a subset $V'\subseteq V$ of variable nodes.
Then, the probability measure on $\mathcal{X}:=\prod_{i\in V}\mathcal{X}_i$ associated with the factor graph $G=(V,F,E,(f_i)_{i\in V}, (f_a)_{a\in F})$ is defined by
\begin{align*}
p(\bm{x}) &= \frac1{Z(G)} \prod_{a\in F} f_a(\bm{x}_{\partial a})\prod_{i\in V} f_i(x_i)\\
Z(G) &:= \sum_{\bm{x}\in\mathcal{X}} \prod_{a\in F} f_a(\bm{x}_{\partial a})\prod_{i\in V} f_i(x_i).
\end{align*}
Here, the constant $Z(G)$ for the normalization is called the partition function, which plays an important role in statistical physics,
information theory, machine learning and computer science~\cite{mezard2009ipa}.

If all degrees of variable nodes are two, a factor graph is called a normal factor graph.
When a set $F$ of factor nodes in a normal factor graph can be separated into two disjoint sets $F_1$ and $F_2$
such that $\partial {a}\cap\partial {a'}=\varnothing$ for $a$ and $a'$ both in $F_1$ or both in $F_2$,
the normal factor graph is said to be bipartite.
Any factor graph can be transformed to an ``equivalent'' bipartite normal factor graph,
by replacing edges by degree-two variable nodes and by replacing original variable nodes by the equality constraints.
For normal factor graphs, a variable can be expressed by an edge since the degrees of variable nodes are two.
Let $(F_1, F_2, E\subseteq F_1\times F_2)$ be a bipartite graph.
For each $(v,w)\in E$, there is an associated finite alphabet $\mathcal{X}_{v,w}$.
For each $v\in F_1$ and $w\in F_2$, there are associated functions $f_v\colon \prod_{w\in\partial v}\mathcal{X}_{v,w}\to\mathbb{R}_{\ge 0}$
and $g_w\colon \prod_{v\in\partial w}\mathcal{X}_{v,w}\to\mathbb{R}_{\ge 0}$, respectively.
A bipartite normal factor graph is denoted by $(F_1, F_2, E\subseteq F_1\times F_2, (f_v)_{v\in F_1}, (g_w)_{w\in F_2})$,
whose partition function is
\begin{align}
Z(G) &:= \sum_{\bm{x}\in\mathcal{X}} \prod_{v\in F_1} f_v(\bm{x}_{\partial v})\prod_{w\in F_2} g_w(\bm{x}_{\partial w})
\label{eq:cm}
\end{align}
where $\mathcal{X}:=\prod_{(v,w)\in E}\mathcal{X}_{v,w}$, 
$\bm{x}_{\partial v}:= (x_{v,w})_{w\in\partial v}$ and
$\bm{x}_{\partial w}:= (x_{v,w})_{v\in\partial w}$ for $v\in F_1$ and $w\in F_2$.

\section{Holographic transformation for bipartite normal factor graphs}
In this section, we briefly introduce the holographic transformation for bipartite normal factor graphs.
Let $\phi_{v,w}\colon \mathcal{X}_{v,w}\times\mathcal{X}_{v,w} \to \mathbb{R}$ and
$\hat{\phi}_{v,w}\colon \mathcal{X}_{v,w}\times\mathcal{X}_{v,w} \to \mathbb{R}$ be mappings for each $(v,w)\in E$ satisfying
\begin{equation*}
\sum_{y\in\mathcal{X}_{v,w}} \phi_{v,w}(x, y)\hat{\phi}_{v,w}(y, z) = \delta(x,z)
\end{equation*}
where $\delta(x,z)$ takes 1 if $x=z$ and 0 otherwise.
Then, it holds
\begin{align*}
&Z(G) = \sum_{\bm{x}\in\mathcal{X}} \prod_{v\in F_1} f_v(\bm{x}_{\partial v})\prod_{w\in F_2} g_w(\bm{x}_{\partial w})\\
&= \sum_{\bm{x}\in\mathcal{X}, \bm{z}\in\mathcal{X}} \prod_{v\in F_1} f_v(\bm{x}_{\partial v})\prod_{w\in F_2} g_w(\bm{z}_{\partial w})
\prod_{(v,w)\in E} \delta(x_{v,w}, z_{v,w})\\
&= \sum_{\bm{x}\in\mathcal{X}, \bm{z}\in\mathcal{X}} \prod_{v\in F_1} f_v(\bm{x}_{\partial v})\prod_{w\in F_2} g_w(\bm{z}_{\partial w})\\
&\qquad \cdot \prod_{(v,w)\in E} \left(\sum_{y\in\mathcal{X}_{v,w}} \phi_{v,w}(x,y)\hat{\phi}_{v,w}(y,z)\right)\\
&= \sum_{\bm{y}\in\mathcal{X}} \prod_{v\in F_1} \left(\sum_{\bm{x}_{\partial v}}f_v(\bm{x}_{\partial v})\prod_{w\in\partial v} \phi_{v,w}(x_{v,w},y_{v,w})\right)\\
&\qquad\cdot\prod_{w\in F_2} \left(\sum_{\bm{z}_{\partial w}} g_w(\bm{z}_{\partial w})\prod_{v\in\partial w}\hat{\phi}_{v,w}(y_{v,w},z_{v,w})\right).
\end{align*}
By letting
\begin{align*}
\hat{f}_v(\bm{y}_{\partial v})&:=\sum_{\bm{x}_{\partial v}}f_v(\bm{x}_{\partial v})\prod_{w\in\partial v} \phi_{v,w}(x_{v,w},y_{v,w})\\
\hat{g}_w(\bm{y}_{\partial w}) &:= \sum_{\bm{z}_{\partial w}} g_w(\bm{z}_{\partial w})\prod_{v\in\partial w}\hat{\phi}_{v,w}(y_{v,w},z_{v,w})
\end{align*}
one obtains
\begin{equation*}
Z(G) = \sum_{\bm{y}\in\mathcal{X}} \prod_{v\in F_1} \hat{f}_v(\bm{y}_{\partial v})\prod_{w\in F_2} \hat{g}_w(\bm{y}_{\partial w}).
\label{eq:cholant}
\end{equation*}
This equality is called the Holant theorem~\cite{valiant2008holographic}, \cite{5695119}, which explains many known equalities~\cite{forney2011partition}.

\section{Bipartite model: Inner product of vectors decomposed to tensor product}
\subsection{Motivation}
The partition function~\eqref{eq:cm} of a bipartite normal factor graph can be regarded as an inner product
of vectors of dimension $|\mathcal{X}|$ in the following way.
Let $\mathcal{V}_{v,w}$ be a linear space on $\mathbb{R}$ of dimension $|\mathcal{X}_{v,w}|$ for $(v,w)\in E$.
Let $(e^{v,w}_x)_{x\in\mathcal{X}_{v,w}}$ be an orthonormal basis of $\mathcal{V}_{v,w}$ for $(v,w)\in E$.
Let $\mathcal{V}_{\partial v}:=\bigotimes_{w\in\partial v}\mathcal{V}_{v,w}$ for $v\in F_1$, 
$\mathcal{V}_{\partial w}:=\bigotimes_{v\in\partial w}\mathcal{V}_{v,w}$ for $w\in F_2$
and $\mathcal{V}:=\bigotimes_{(v,w)\in E}\mathcal{V}_{v,w}$
where $\otimes$ denotes the tensor product.
Let $\mathrm{f}_v$ be a vector in $\mathcal{V}_{\partial v}$ defined by
\begin{equation*}
\mathrm{f}_v := \sum_{\bm{x}_{\partial v}\in\prod_{w\in\partial v}\mathcal{X}_{v,w}} f_v(\bm{x}_{\partial v}) \bigotimes_{w\in\partial v} e^{v,w}_{x_{v,w}}.
\end{equation*}
The vector $\mathrm{g}_w\in\mathcal{V}_{\partial w}$ is also defined in the same way.
Here,
it holds
\begin{align*}
\bigotimes_{v\in F_1} \mathrm{f}_v &= \sum_{\bm{x}\in\mathcal{X}} \prod_{v\in F_1}f_v(\bm{x}_{\partial v}) \bigotimes_{(v,w)\in E} e^{v,w}_{x_{v,w}}\\
\bigotimes_{w\in F_2} \mathrm{g}_w &= \sum_{\bm{x}\in\mathcal{X}} \prod_{w\in F_2}g_w(\bm{x}_{\partial w}) \bigotimes_{(v,w)\in E} e^{v,w}_{x_{v,w}}.
\end{align*}
Hence, the partition function~\eqref{eq:cm} of bipartite normal factor graph is equal to
\begin{equation*}
\left\langle \bigotimes_{v\in F_1} \mathrm{f}_v, \bigotimes_{w\in F_2} \mathrm{g}_w\right\rangle
\end{equation*}
where $\langle\cdot,\cdot\rangle$ denotes the inner product of vectors.
In this paper, the holographic transformation is dealt with on this representation of the partition function.
Since the holographic transformation can be defined for the partition function on any field $\mathbb{F}$,
and since an inner product can be defined only for linear spaces on $\mathbb{R}$ or $\mathbb{C}$,
we have to consider linear spaces on a general field $\mathbb{F}$ and bilinear forms instead of inner products.

\subsection{Bilinear form, adjoint map and tensor product}
Let $\mathbb{F}$ be a field.
Let $\mathcal{V}$ be a $q$-dimensional linear space on $\mathbb{F}$.
Let $\langle \cdot, \cdot \rangle_{\mathcal{V}}\colon \mathcal{V}\times \mathcal{V}\to \mathbb{F}$ be a bilinear form.
Let $(e_x)_{x=0,\dotsc,q-1}$ be an arbitrarily chosen basis for $\mathcal{V}$.
A bilinear form is represented by the $q\times q$ coefficient matrix $K_{\mathcal{V}}$ whose $(x,{x'})$ element is $\langle e_x, e_{x'}\rangle_{\mathcal{V}}$
for $x,x'\in\{0,\dotsc,q-1\}$.
Then, the bilinear form is expressed by
$\langle f, g \rangle_{\mathcal{V}} = f^T K_{\mathcal{V}} g$
where $f,g\in\mathcal{V}$ are represented by vectors with respect to the basis $(e_x)_x$.
In this paper, we always assume that the bilinear form is non-degenerate, i.e., the coefficient matrix $K_{\mathcal{V}}$ is invertible.
Let $\mathcal{V}$ and $\mathcal{W}$ be linear spaces with non-degenerate bilinear forms 
$\langle\cdot,\cdot\rangle_{\mathcal{V}}$ and $\langle\cdot,\cdot\rangle_{\mathcal{W}}$, 
with coefficient matrices $K_{\mathcal{V}}$ and $K_{\mathcal{W}}$, respectively.
It holds for any $f\in \mathcal{V}$, $g\in \mathcal{W}$ and a linear map $A\colon \mathcal{V}\to\mathcal{W}$ that
$\langle A f, g \rangle_{\mathcal{W}}=
\langle f, A^* g \rangle_{\mathcal{V}}$
where $A^*=K_{\mathcal{V}}^{-1} A^T K_{\mathcal{W}}$ is called the right adjoint map of $A$.

Let $\mathcal{V}$ and $\mathcal{W}$ be linear spaces on $\mathbb{F}$ with bilinear forms $\langle\cdot,\cdot\rangle_{\mathcal{V}}$,
$\langle\cdot,\cdot\rangle_{\mathcal{W}}$, respectively.
A bilinear form $\langle \cdot,\cdot\rangle_{\mathcal{V}\otimes\mathcal{W}}$ for the tensor product space $\mathcal{V}\otimes\mathcal{W}$
is defined by
\begin{equation*}
\langle f_1\otimes g_1, f_2\otimes g_2 \rangle_{\mathcal{V}\otimes\mathcal{W}} =
\langle f_1, f_2 \rangle_{\mathcal{V}} \langle g_1, g_2 \rangle_{\mathcal{W}}
\end{equation*}
for any $f_1, f_2\in\mathcal{V}$ and $g_1, g_2\in\mathcal{W}$.
It is easy to check that $(A_{\mathcal{V}}\otimes A_{\mathcal{W}})^*=A_{\mathcal{V}}^*\otimes A_{\mathcal{W}}^*$
where $A_{\mathcal{V}}$ and $A_{\mathcal{W}}$ are arbitrary linear maps on $\mathcal{V}$ and $\mathcal{W}$, respectively.

\subsection{Bipartite model and holographic transformation}
Let $(V, W, E\subseteq V\times W)$ be a bipartite graph.
Let $\mathbb{F}$ be a field.
For each edge $(v,w)\in E$, there is an associated linear space $\mathcal{V}_{v,w}$ of dimension $q_{v,w}$ with a non-degenerate 
bilinear form $\langle\cdot,\cdot\rangle_{\mathcal{V}_{v,w}}$.
Let $\mathcal{V}_{\partial v}:=\bigotimes_{w\in\partial v}\mathcal{V}_{v,w}$, 
$\mathcal{V}_{\partial w}:=\bigotimes_{v\in\partial w}\mathcal{V}_{v,w}$
and $\mathcal{V}:=\bigotimes_{(v,w)\in E}\mathcal{V}_{v,w}$.
For each $v\in V$, there is a vector $f_v\in\mathcal{V}_{\partial v}$.
Similarly, for each $w\in W$, there is a vector $g_w\in\mathcal{V}_{\partial w}$.
The ``partition function'' of the bipartite model $(V, W, E, (f_v)_{v\in V}, (g_w)_{w\in W})$ is defined by
\begin{equation}
\left\langle \bigotimes_{v\in V} f_v, \bigotimes_{w\in W} g_w\right\rangle_{\mathcal{V}}.
\label{eq:inner}
\end{equation}

The partition function of a bipartite normal factor graph~\eqref{eq:cm} can be expressed in this form.
Conversely, \eqref{eq:inner} can be represented by the partition function of a bipartite normal factor graph on 
the field $\mathbb{F}$ without the non-negative condition on weight as follows.
Let $(e^{v,w}_x)_{x=0,\dotsc,q_{v,w}-1}$ be an orthonormal basis of $\mathcal{V}_{v,w}$ for $(v,w)\in E$.
Then, \eqref{eq:inner} can be expanded into the form
\begin{align}
&\sum_{\bm{x}\in\mathcal{X}}\left\langle\bigotimes_{v\in V} f_v, \bigotimes_{(v,w)\in E} e_{x_{v,w}}^{v,w} \right\rangle_{\mathcal{V}}
\left\langle \bigotimes_{(v,w)\in E} e_{x_{v,w}}^{v,w}, \bigotimes_{w\in W} g_w \right\rangle_{\mathcal{V}}\nonumber\\
&=\sum_{\bm{x}\in\mathcal{X}}\prod_{v\in V}\left\langle f_v, \bigotimes_{w\in\partial v} e_{x_{v,w}}^{v,w} \right\rangle_{\mathcal{V}_{\partial v}}
\prod_{w\in W} \left\langle \bigotimes_{v\in\partial w} e_{x_{v,w}}^{v,w}, g_w \right\rangle_{\mathcal{V}_{\partial w}}.
\label{eq:expand}
\end{align}
Hence, the bipartite model does not have much power of representation than the bipartite normal factor graph.
However, the expansion with respect to some particular basis is not necessarily natural.
The bilinear form interpretation gives a simple derivation of the Holant theorem as follows.

\begin{theorem}[Holant theorem for the bipartite model]
Let $\Phi_{v,w}$ be an invertible linear map on $\mathcal{V}_{v,w}$ and $\hat{\Phi}_{v,w}$ be the inverse map of $\Phi_{v,w}$
for $(v,w)\in E$.
Then, it holds
\begin{align*}
\left\langle \bigotimes_{v\in V} f_v, \bigotimes_{w\in W} g_w\right\rangle_{\mathcal{V}}
=
\left\langle \bigotimes_{v\in V} \hat{f}_v, \bigotimes_{w\in W} \hat{g}_w\right\rangle_{\mathcal{V}}
\end{align*}
where
\begin{align*}
\hat{f}_v &= \left(\bigotimes_{w\in\partial v} \hat{\Phi}_{v,w}\right)(f_v),&
\hat{g}_w &= \left(\bigotimes_{v\in\partial w} \Phi^*_{v,w}\right)(g_w).
\end{align*}
\end{theorem}
\begin{IEEEproof}
\begin{align*}
&\left\langle \bigotimes_{v\in V} f_v, \bigotimes_{w\in W} g_w\right\rangle_{\mathcal{V}}\\
&=
\left\langle \left(\bigotimes_{(v,w)\in E} \Phi_{v,w} \circ \hat{\Phi}_{v,w}\right)\left(\bigotimes_{v\in V} f_v\right), \bigotimes_{w\in W} g_w\right\rangle_{\mathcal{V}}\\
&=
\Biggl\langle \left(\bigotimes_{(v,w)\in E} \hat{\Phi}_{v,w}\right)\left(\bigotimes_{v\in V} f_v\right),\\
&\quad\left(\bigotimes_{(v,w)\in E} \Phi^*_{v,w}\right)\left(\bigotimes_{w\in W} g_w\right)\Biggr\rangle_{\mathcal{V}}
=
\left\langle \bigotimes_{v\in V} \hat{f}_v, \bigotimes_{w\in W} \hat{g}_w\right\rangle_{\mathcal{V}}.
\end{align*}
\end{IEEEproof}

\subsection{Belief propagation for the bipartite model}\label{sec:bp}
In this subsection, we assume $\mathbb{F}=\mathbb{R}$ for defining the belief propagation on the bipartite model.
The bilinear forms are assumed to be inner products.
Let $C_{v,w}$ be a closed convex cone in the inner product space $\mathcal{V}_{v,w}$ for $(v,w)\in E$.
For a closed convex cone $C$ in an inner product space $\mathcal{V}$, the dual cone of $C$ is denoted by $C^*$, i.e.,
$C^*:=\{f\in\mathcal{V}\mid \langle f, g\rangle_{\mathcal{V}}\ge 0\quad \forall g\in C\}$.
Let $\bigotimes_{v\in \partial w} C_{v,w}$ be a closed convex cone generated by $\{\bigotimes_{v\in\partial w}F_{v}\mid F_v\in C_{v,w}\}$.
Let $C_{v}:=(\bigotimes_{w\in\partial v} C^*_{v,w})^*$ and
$C^*_{w}:=(\bigotimes_{v\in\partial w} C_{v,w})^*$.
We assume that $f_v\in C_{v}$ and $g_w\in C^*_{w}$.
Let $f$ and $g$ be vectors in inner product spaces $\mathcal{V}\otimes\mathcal{W}$ and $\mathcal{V}$, respectively.
Then, the partial inner product $\langle f,g\rangle_{\mathcal{V}}\in\mathcal{W}$ is defined as the unique vector satisfying
$\langle \langle f,g\rangle_{\mathcal{V}}, h\rangle_{\mathcal{W}}=
\langle f,g\otimes h\rangle_{\mathcal{V}\otimes\mathcal{W}}$ for any $h\in\mathcal{W}$.
Let $\mathcal{V}_{\partial v\setminus w}:=\bigotimes_{w'\in\partial v\setminus \{w\}}\mathcal{V}_{v,w'}$ and
$\mathcal{V}_{\partial w\setminus v}:=\bigotimes_{v'\in\partial w\setminus \{v\}}\mathcal{V}_{v',w}$ for $(v,w)\in E$.
\begin{definition}[Belief propagation]
Let $u_{v,w}$ and $u^*_{v,w}$ be vectors in interiors of $C_{v,w}$ and $C^*_{v,w}$, respectively.
Let $(m_{v\to w}^{(0)}\in C_{v,w})_{(v,w)\in E}$ be arbitrarily chosen initial messages.
Then, in the belief propagation, the messages are updated according to the following rules
\begin{align*}
m^{(t)}_{v\to w} &= \frac1{Z^{(t)}_{v\to w}}\left\langle f_v,\, \bigotimes_{w'\in\partial v\setminus\{w\}} m^{(t)}_{w'\to v}\right\rangle_{\mathcal{V}_{\partial v\setminus w}}\\
m^{(t)}_{w\to v} &= \frac1{Z^{(t)}_{w\to v}}\left\langle \bigotimes_{v'\in\partial w\setminus\{v\}} m^{(t-1)}_{v'\to w},\, g_w\right\rangle_{\mathcal{V}_{\partial w\setminus v}}
\end{align*}
for $t=1,2,\dotsc$ for all $(v,w)\in E$
where the strictly positive constants $Z^{(t)}_{v\to w}$ and $Z^{(t)}_{w\to v}$ are chosen such that $\langle m^{(t)}_{v\to w}, u^*_{v,w}\rangle_{\mathcal{V}_{v,w}} = 1$
and $\langle  u_{v,w}, m^{(t)}_{w\to v}\rangle_{\mathcal{V}_{v,w}} =1$, respectively.
\end{definition}
Note that it always holds $m^{(t)}_{v\to w}\in C_{v,w}$ and $m^{(t)}_{w\to v}\in C^*_{v,w}$
for $(v,w)\in E$.
\begin{remark}
For the standard belief propagation for bipartite normal factor graphs,
 the closed convex cone $C_{v,w}$ corresponds to the set of non-negative vectors,
which is self-dual, i.e., $C^*_{v,w}=C_{v,w}$.
The vectors $u_{v,w}$ and $u^*_{v,w}$ correspond to the all-one vector.
\end{remark}
Similarly to the standard belief propagation for factor graphs, 
the general belief propagation defined above gives the exact computation of the partition function
of bipartite models on a cycle-free bipartite graph.
It will be shown as a corollary of the loop calculus formula in the next section.

\subsection{Loop calculus for the bipartite model}\label{sec:loop}
In this section, we derive the loop calculus formula for the bipartite model.
The derivation in this section is essentially equivalent to that in~\cite{mori2015lcn}.
Let $(e^{v,w}_x)_{x=0,\dotsc,q_{v,w}-1}$ be an arbitrarily chosen orthonormal basis for $\mathcal{V}_{v,w}$ for $(v,w)\in E$.
For the loop calculus, the following additional conditions are required
\begin{equation}
\begin{split}
\left\langle \hat{f}_v,\, e^{v,w}_x\otimes \bigotimes_{w'\in\partial v\setminus \{w\}} e^{v,w'}_0\right\rangle_{\mathcal{V}_{\partial v}}&=0\\
\left\langle e^{v,w}_x\otimes \bigotimes_{v'\in\partial w\setminus \{v\}} e^{v',w}_0,\,\hat{g}_w\right\rangle_{\mathcal{V}_{\partial w}}&=0
\end{split}
\label{eq:cond}
\end{equation}
for any $(v,w)\in E$ and $x\in \{1,\dotsc,q_{v,w}-1\}$.
These conditions are equivalent to
\begin{align*}
\left\langle f_v,\, \hat{\Phi}^*_{v,w}(e^{v,w}_x)\otimes \bigotimes_{w'\in\partial v\setminus \{w\}} \hat{\Phi}^*_{v,w'}(e^{v,w'}_0)\right\rangle_{\mathcal{V}_{\partial v}}&=0\\
\left\langle \Phi_{v,w}(e^{v,w}_x)\otimes \bigotimes_{v'\in\partial w\setminus \{v\}} \Phi_{v,w}(e^{v',w}_0),\,g_w\right\rangle_{\mathcal{V}_{\partial w}}&=0
\end{align*}
for any $(v,w)\in E$ and $x\in \{1,\dotsc,q_{v,w}-1\}$.
Furthermore, these conditions are equivalent to
\begin{align*}
\left\langle \left\langle f_v,\, \bigotimes_{w'\in\partial v\setminus \{w\}} \hat{\Phi}^*_{v,w'}(e^{v,w'}_0)\right\rangle_{\mathcal{V}_{\partial v\setminus w}},\,\hat{\Phi}^*_{v,w}(e^{v,w}_x) \right\rangle_{\mathcal{V}_{v,w}}&=0\\
\left\langle \Phi_{v,w}(e^{v,w}_x),\, \left\langle \bigotimes_{v'\in\partial w\setminus \{v\}} \Phi_{v,w}(e^{v',w}_0),\,g_w\right\rangle_{\mathcal{V}_{\partial w\setminus v}}\right\rangle_{\mathcal{V}_{v,w}}&=0
\end{align*}
for any $(v,w)\in E$ and $x\in \{1,\dotsc,q_{v,w}-1\}$.
Since $\hat{\Phi}_{v,w}(\Phi_{v,w}(e^{v,w}_x))=e^{v,w}_x$,
it holds 
\begin{align}
 \delta(x, x')&=
\left\langle \hat{\Phi}_{v,w}(\Phi_{v,w}(e^{v,w}_x)),\,e^{v,w}_{x'}\right\rangle_{\mathcal{V}_{v,w}}\nonumber\\
 &= \left\langle \Phi_{v,w}(e^{v,w}_x),\, \hat{\Phi}^*_{v,w}(e^{v,w}_{x'})\right\rangle_{\mathcal{V}_{v,w}}
\label{eq:id}
\end{align}
for any $x,x'\in\{0,1,\dotsc,q_{v,w}-1\}$ and $(v,w)\in E$.
Hence,
\begin{align*}
\left\langle f_v,\, \bigotimes_{w'\in\partial v\setminus \{w\}} \hat{\Phi}^*_{v,w'}(e^{v,w'}_0)\right\rangle_{\mathcal{V}_{\partial v\setminus w}}
=
\alpha_{v,w} \Phi_{v,w}(e^{v,w}_0)\\
\left\langle \bigotimes_{v'\in\partial w\setminus \{v\}} \Phi_{v,w}(e^{v',w}_0),\,g_w\right\rangle_{\mathcal{V}_{\partial w\setminus v}}
=
\hat{\alpha}_{v,w} \hat{\Phi}^*_{v,w}(e^{v,w}_0)
\end{align*}
for any $(v,w)\in E$ 
where 
$\alpha_{v,w}:=
 \bigl\langle \hat{f}_v,\allowbreak\, \bigotimes_{w\in\partial v}e^{v,w}_0\bigr\rangle_{\mathcal{V}_{\partial v}}$ and 
$\hat{\alpha}_{v,w}:=
\bigl\langle \bigotimes_{v\in\partial w} e^{v,w}_0,\allowbreak\, \hat{g}_w\bigr\rangle_{\mathcal{V}_{\partial w}}$.
Furthermore, we assume that $\Phi_{v,w}(e_0^{v,w})\in C_{v,w}$ and
$\hat{\Phi}^*_{v,w}(e_0^{v,w})\in C^*_{v,w}$ for $(v,w)\in E$.
Then, without loss of generality,
it holds
$\Phi_{v,w}(e^{v,w}_0)=c_{v,w}m_{v\to w}$  and
$\hat{\Phi}^*_{v,w}(e^{v,w}_0)=\hat{c}_{v,w}m_{w\to v}$
for some strictly positive constants $c_{v,w}, \hat{c}_{v,w}\in\mathbb{R}_{>0}$ and
for some $m_{v\to w}\in C_{v,w}$ and $m_{w\to v}\in C^*_{v,w}$
satisfying $\langle m_{v\to w}, u^*_{v,w}\rangle_{\mathcal{V}_{v,w}}=1$ and
$\langle u_{v,w}, m_{w\to v}\rangle_{\mathcal{V}_{v,w}}=1$, respectively.
Then, it is easy to check that
$(m_{v\to w})_{(v,w)\in E}$ and $(m_{w\to v})_{(v,w)\in E}$ have to satisfy the fixed point equation of the belief propagation
and
$c_{v,w}\hat{c}_{v,w}=\langle m_{v\to w}, m_{w\to v}\rangle_{\mathcal{V}_{v,w}}^{-1}$.

In the expansion~\eqref{eq:expand} for the new expression of the inner product with respect to the orthonormal basis,
the weight of the all-zero assignment is
\begin{align*}
&\prod_{v\in V} \left\langle f_v,\, \bigotimes_{w\in\partial v} m_{w\to v} \right\rangle_{\mathcal{V}_{\partial v}}
\prod_{w\in W} \left\langle \bigotimes_{v\in\partial w} m_{v\to w},\, g_w \right\rangle_{\mathcal{V}_{\partial w}}\\
&\quad \cdot \prod_{(v,w)\in E} \frac1{\left\langle m_{v\to w}, m_{w\to v}\right\rangle_{\mathcal{V}_{v,w}}}.
\end{align*}
This quantity can be regarded as the ``Bethe approximation'' of the inner product~\eqref{eq:inner} on the chosen fixed point of the belief propagation~\cite{mori2015lcn}.
In the summation~\eqref{eq:expand}, only assignments $\bm{x}\in\mathcal{X}$ whose non-zero part corresponds to some generalized loop have non-zero weight
due to the conditions~\eqref{eq:cond}~\cite{mori2015lcn}.
Hence, if the bipartite graph is cycle-free, the Bethe approximation is exactly equal to the inner product~\eqref{eq:inner}.
Choices of $(\Phi_{v,w}(e_x^{v,w}))_{x=1,\dotsc,q_{v,w}-1}$ and
$(\hat{\Phi}^*_{v,w}(e_x^{v,w}))_{x=1,\dotsc,q_{v,w}-1}$ are arbitrary if they satisfy~\eqref{eq:id} for any $(v,w)\in E$.
For bipartite normal factor graphs, expressions of the remaining degree of freedom using ideas from information geometry was shown in~\cite{mori2015lcn}.

\section{Bipartite quantum model}
\subsection{Inner product space spanned by Hermitian matrices}
In this section, we consider the inner product space spanned by Hermitian matrices
for considering problems in quantum information science.
Let $\mathcal{H}$ be an inner-product space on $\mathbb{C}$.
The inner product for the inner-product space $\mathcal{H}$ is denoted by $\langle\cdot,\cdot\rangle_{\mathcal{H}}$.
A linear operator $A\colon\mathcal{H}\to\mathcal{H}$ satisfying $\langle A \psi,\phi\rangle_{\mathcal{H}} = \langle \psi, A \phi\rangle_{\mathcal{H}}$ for any $\psi,\phi\in\mathcal{H}$
is called an Hermitian linear operator.
A set $\mathcal{L}_{\mathrm{h}}(\mathcal{H})$ of Hermitian linear operators acting on $\mathcal{H}$
can be regarded as an inner-product space on $\mathbb{R}$ with the conventional addition, the conventional scalar multiplication and 
the Hilbert-Schmidt inner product $\langle A, B\rangle_{\mathcal{L}_{\mathrm{h}}(\mathcal{H})} := \Tr(A B)$
for $A, B\in\mathcal{L}_{\mathrm{h}}(\mathcal{H})$.
Note that the dimension of $\mathcal{L}_{\mathrm{h}}(\mathcal{H})$ is square of the dimension of $\mathcal{H}$.

\subsection{Bipartite quantum model}
Let $(V, W, E\subseteq V\times W)$ be a bipartite graph.
For each edge $(v, w)\in E$, there is an associated inner product space $\mathcal{H}_{v,w}$ on $\mathbb{C}$.
For each $v\in V$ and $w\in W$, there are associated positive-semidefinite Hermitian operators 
$f_{v}$ on $\bigotimes_{w\in\partial v} \mathcal{H}_{v,w}$ and
$g_{w}$ on $\bigotimes_{v\in\partial w} \mathcal{H}_{v,w}$,
 respectively.
Then, a bipartite quantum model is denoted by $(V, W, E, (f_v)_{v\in V}, (g_w)_{w\in W})$
which represents a non-negative value
\begin{equation}
\Tr\left(\bigotimes_{v\in V} f_v \bigotimes_{w\in W} g_w\right).
\label{eq:qm}
\end{equation}
The bipartite quantum model can be regarded as special cases of the bipartite model
by letting $\mathcal{V}_{v,w}=\mathcal{L}_{\mathrm{h}}(\mathcal{H}_{v,w})$ and by using the Hilbert-Schmidt inner product for $(v,w)\in E$.
For the belief propagation,
the closed convex cone $C_{v,w}$ for $(v,w)\in E$ corresponds to the set of positive-semidefinite matrices on $\mathcal{H}_{v,w}$,
which is a self-dual convex cone with respect to the Hilbert-Schmidt inner product.
For the Hermitian matrices $u_{v,w}$ and $u^*_{v,w}$, one can choose the identity matrix.
Note that $\bigotimes_{v\in\partial w} C_{v,w}$ is a proper subset of the set of positive-semidefinite matrices on 
$\bigotimes_{v\in\partial w}\mathcal{H}_{v,w}$.
Hence, its dual $\left(\bigotimes_{v\in\partial w} C_{v,w}\right)^*$ is a proper superset of
the set of positive-semidefinite matrices on $\bigotimes_{v\in\partial w}\mathcal{H}_{v,w}$~\cite{Ando20043}.

\subsection{Applications}
There are several applications of the bipartite quantum model~\eqref{eq:qm} as follows.
Let $\mathcal{H}$ be a Hilbert space.
Let $\rho$ be a quantum state on $\mathcal{H}$.
Let $(P, I-P)$ be a positive operator-valued measure (POVM) on $\mathcal{H}$.
A probability of the outcome corresponding to $P$ of the POVM for the quantum state $\rho$ is $\Tr(\rho P)$.
The value $\Tr(\rho P)$ can be represented by a bipartite quantum model $(V=\{v\}, W=\{w\}, E=\{(v,w)\}, (f_{v}=\rho), (g_{w}=P))$.
If $\hat{\Phi}_{v,w}$ is a trace-preserving completely-positive map, the Holant theorem corresponds to the equivalence
of the Schr\"{o}dinger picture and the Heisenberg picture~\cite{wolf2012quantum}.
As another application, we can deal with the quantum teleportation-type problems
$(V=\{v_1,v_2\}, W=\{w_1,w_2\}, E=\{(v_1,w_1),\allowbreak (v_1,w_2), (v_2, w_2)\}, 
(f_{v_1}=\tau, f_{v_2}=\rho),\allowbreak
 (g_{w_1}=P,\allowbreak g_{w_2}=Q))$
where $\tau\in\mathcal{L}_{\mathrm{h}}(\mathcal{H}_{v_1,w_1}\otimes\mathcal{H}_{v_1,w_2})$
is an arbitrary entangled quantum state, and where
$Q\in\mathcal{L}_{\mathrm{h}}(\mathcal{H}_{v_1,w_2}\otimes\mathcal{H}_{v_2,w_2})$ is a positive semidefinite operator
corresponding to outcome of some POVM\@.
Here, $\mathcal{H}_{v_1,w_1}$ and $\mathcal{H}_{v_1,w_2}\otimes\mathcal{H}_{v_2,w_2}$ correspond to Bob's Hilbert space
and Alice's Hilbert space, respectively.

More generally, we can deal with larger systems including cycles.
As an important example, we introduce the measurement probability in the measurement-based quantum computation (MBQC). 
In MBQC, the probability that the outcomes of the measurements are $(\gamma_i)_{i\in V}$ is denoted by
$|(\bigotimes_{i\in V} \bra{\gamma_i})\ket{G}|^2$ where $\ket{G}\in\bigotimes_{i\in V}\mathcal{H}_i$ 
is a graph state represented by a graph $G=(V,E)$, which is a special stabilizer state~\cite{PhysRevA.68.022312}.
The projected entangled-pair state representation of graph states
is
$\ket{G}=2^{|E|-|V|/2}\bigotimes_{i\in V} P_i \bigotimes_{\{i,j\}\in E} \ket{\Omega}_{i,j}$
where $P_i:=\ket{0}_i\bigotimes_{j\in\partial i}\bra{0}_{i,j} + \ket{1}_i\bigotimes_{j\in\partial i}\bra{1}_{i,j}$ 
and $\ket{\Omega}_{i,j}=U_{\mathrm{CZ}}\ket{+}_{i,j}\ket{+}_{j,i}$~\cite{PhysRevA.70.060302}.
Here, $U_{\mathrm{CZ}}$ and $\ket{+}$ are the controlled-Z operation and $(\ket{0}+\ket{1})/\sqrt{2}$, respectively.
Hence, the probability $|(\bigotimes_{i\in V} \bra{\gamma_i})\ket{G}|^2=\Tr((\bigotimes_{i\in V}\ket{\gamma_i}\bra{\gamma_i})\ket{G}\bra{G})$
 can be written as
$2^{2|E|-|V|}$ times
\begin{equation}
\Tr\left( \bigotimes_{i\in V} \left(P_i^\dagger \ket{\gamma_i}\bra{\gamma_i}P_i\right) \bigotimes_{\{i,j\}\in E} \Ket{\Omega_{\{i,j\}}}\Bra{\Omega_{\{i,j\}}}\right)
\label{eq:mbqc}
\end{equation}
which is an instance of the bipartite quantum model~\eqref{eq:qm}.
If~\eqref{eq:mbqc} can be accurately approximated by the classical computation efficiently when $G$ is the two-dimensional square lattice,
any quantum computation can be efficiently simulated by the classical computer~\cite{PhysRevA.68.022312}.
Hence, \eqref{eq:mbqc} is an important quantity although, on the other hand, 
we believe that the classical computer cannot efficiently simulate the universal quantum computation.
From this work, the holographic transformation, belief propagation and loop calculus can be applied for~\eqref{eq:mbqc}.
Belief propagations for problems in quantum physics were suggested in~\cite{PhysRevB.76.201102}, \cite{Leifer2008quantum}, \cite{PhysRevB.78.134424}.
The belief propagation suggested in Section~\ref{sec:bp} for the bipartite quantum model is novel
and is derived only by using linear algebra as in Section~\ref{sec:loop} in contrast to the other quantum belief propagations.

\section{The generalized probabilistic theories}
The generalized probabilistic theories (GPT) are general theories of probabilities including the classical probability theory and the
quantum probability thoery~\cite{1751-8121-47-32-323001}.
In GPT, a state is an element of a cone $C$ in a linear space.
An effect, which means a measurement, is an element of the dual cone $C^*$.
A probability of outcome corresponding to $e\in C^*$ on a state $\omega\in C$ is 
the inner product $\langle e,\omega\rangle$.
Hence, 
the probability of outcome corresponding to $\bigotimes_{w\in W}g_{w}$ on a state $\bigotimes_{v\in V}f_v$ is written as~\eqref{eq:inner}.
The belief propagation and loop calculus are naturally generalized to the measurement probability in the GPT\@.

\section*{Acknowledgment}
This work was supported by MEXT KAKENHI Grant Number 24106008.
The author would like to thank Keisuke~Fujii for teaching me basics of the quantum computation. 
The author would also like to thank Tomoyuki~Morimae for letting me know the generalized probabilistic theories.

\bibliographystyle{IEEEtran}
\bibliography{IEEEabrv,biblio}

\newpage
\appendix
\section{Loop calculus for the bipartite quantum model}\label{sec:ql}
In this appendix, the expressions of $(\Phi_{v,w}(e_x^{v,w}))_{x=1,\dotsc,q_{v,w}-1}$ and $(\hat{\Phi}^*_{v,w}(e_x^{v,w}))_{x=1,\dotsc,q_{v,w}-1}$
 for the loop calculus on the bipartite quantum model are suggested.
For the expression, concepts from quantum information geometry are used similarly to the classical case~\cite{mori2015lcn}.
\begin{definition}[Quantum exponential family]
Let $d$ be a positive integer.
Let $\mathcal{H}$ be an inner product space on $\mathbb{C}$.
Let $T_1,\dotsc,T_d$ be Hermitian operators on $\mathcal{H}$.
Assume that $T_1,\dotsc,T_d,I$ are linearly independent where $I$ denotes the identity operator on $\mathcal{H}$.
Then, the quantum exponential family is a parametrized family of density operators defined by
\begin{equation*}
\rho_{\bm{\theta}} = \exp\left\{\sum_{k=1}^d \theta_k T_k - \psi(\bm{\theta}) I\right\}
\end{equation*}
where
\begin{equation*}
\psi(\bm{\theta}) := \log\Tr\left(\exp\left\{\sum_{k=1}^d \theta_k T_k \right\}\right).
\end{equation*}
The parameter $(\theta_1,\dotsc,\theta_d)\in\mathbb{R}^d$ is called a natural parameter.
\end{definition}
\begin{example}\label{exm:qe}
Let $\mathcal{H}$ be an inner product space on $\mathbb{C}$ of dimension $q$.
The family $\{\rho\in\mathcal{L}_{\mathrm{h}}(\mathcal{H})\mid \rho > 0, \Tr\rho=1\}$
can be regarded as $q^2-1$ dimensional exponential family.
\end{example}
It holds
$\frac{\partial \psi(\bm{\theta})}{\partial \theta_k} = \Tr(\rho_{\bm{\theta}} T_k) =: \eta_k$.
Here, $(\eta_1,\dotsc,\eta_d)$ gives another coordinate system for the parametrized family.
The parameter $(\eta_1,\dotsc,\eta_d)$ is called an expectation parameter.
In quantum information geometry, the Bogoliubov-Kubo-Mori metric~\cite{amari2000method} on quantum information manifolds satisfies
\begin{align}
\left\langle \frac{\partial }{\partial \theta_k}, \frac{\partial }{\partial \eta_l}\right\rangle_{\mathrm{BKM}(\rho)}&:=
\int_0^1\Tr\left(\rho^\lambda \frac{\partial \log \rho}{\partial \theta_k}\rho^{1-\lambda} \frac{\partial \log \rho}{\partial \eta_l}\right)\mathrm{d}\lambda\nonumber\\
&=\Tr\left(\frac{\partial \rho}{\partial \eta_l} \frac{\log \rho}{\partial \theta_k}\right)
=
\delta(k,l).
\label{eq:BKM}
\end{align}
Similarly to the classical case~\cite{mori2015lcn}, this property is useful for expressing the remaining degree of freedom in the loop calculus
for the quantum model.

Let $A\star B:=A^{1/2}BA^{1/2}$ for a positive-semidefinite matrix $A$ and an Hermitian matrix $B$.
Let $b_{v,w} := (m_{v\to w} \star m_{w\to v})/\Tr(m_{v\to w} m_{w\to v})$ and
$\bar{b}_{v,w} := (m_{w\to v} \star m_{v\to w})/\Tr(m_{v\to w} m_{w\to v})$ for $(v,w)\in E$.
Assume that $b_{v,w}$ and $\bar{b}_{v,w}$ are positive definite and are regarded as members of 
the quantum exponential family in Example~\ref{exm:qe}.
Fix $c_{v,w}=1$ and
\begin{align*}
\hat{\Phi}^*_{v,w}(e^{v,w}_x)&=m_{v\to w}^{-1} \star \left(\frac{\partial \bar{b}_{v,w}}{\partial \eta_x}\right)\\
&=m_{w\to v}\star \left(m_{w\to v}^{-1}\star\left(m_{v\to w}^{-1} \star \left(\frac{\partial \bar{b}_{v,w}}{\partial \eta_x}\right)\right)\right)\\
\Phi_{v,w}(e^{v,w}_x)&=m_{v\to w} \star \left(\frac{\partial \log b_{v,w}}{\partial \theta_x} \right)
\end{align*}
for $x=1,\dotsc,q_{v,w}-1$ and $(v,w)\in E$.
The above choices satisfy the condition~\eqref{eq:id}.
This fact seems to be similar to the equation~\eqref{eq:BKM} although we deal with two density matrices $b_{v,w}$ and $\bar{b}_{v,w}$
for each $(v,w)\in E$
in contrast to the classical case~\cite{mori2015lcn}.
Conversely, the above choices cover all possible choices of
$(\Phi_{v,w}(e_x^{v,w}))_{x=1,\dotsc,q_{v,w}-1}$ and $(\hat{\Phi}^*_{v,w}(e_x^{v,w}))_{x=1,\dotsc,q_{v,w}-1}$
satisfying~\eqref{eq:id}
since both of them have the degrees of freedom represented by a $(q_{v,w}-1)\times (q_{v,w}-1)$ invertible matrix
for $(v,w)\in E$.
In the above expression,
\begin{equation*}
m_{w\to v}^{-1}\star\left(m_{v\to w}^{-1} \star \left(\frac{\partial \bar{b}_{v,w}}{\partial \eta_x}\right)\right)
\end{equation*}
may be regarded as a variant of logarithmic derivatives of $\bar{b}_{v,w}$.
Note that $\frac{\partial \rho_{\bm{\eta}}}{\partial \eta_x}$ is independent of $\bm{\eta}$.
Especially, if the sufficient statistics $(T_x)_{x=1,\dotsc,q^2-1}$ are chosen such that $(T_1,\dotsc,T_{q^2-1}, (1/\sqrt{q}) I)$
 is an orthonormal basis for $\mathcal{L}_\mathrm{h}(\mathcal{H})$, it holds
$\frac{\partial \rho_{\bm{\eta}}}{\partial \eta_x}=T_x$.
On the other hand, it always holds $\frac{\partial \log\rho_{\bm{\theta}}}{\partial \theta_x}=T_x-\eta_x I$.
By summarizing the above, one obtains
\begin{align*}
\hat{\Phi}^*_{v,w}(e^{v,w}_0)&=\frac1{\langle m_{v\to w}, m_{w\to v}\rangle_{\mathcal{V}_{v,w}}}m_{w\to v}\\
\Phi_{v,w}(e^{v,w}_0)&=m_{v\to w} \\
\hat{\Phi}^*_{v,w}(e^{v,w}_x)&=m_{v\to w}^{-1} \star T_x\\
\Phi_{v,w}(e^{v,w}_x)&=m_{v\to w} \star \left(T_x - \Tr(b_{v,w} T_x) I\right)
\end{align*}
for $x=1,\dotsc,q_{v,w}-1$ and $(v,w)\in E$.
Some results in~\cite{mori2015lcn} for classical factor graphs can be generalized to the bipartite quantum model as well.

\end{document}